\documentclass[aps,prd,amsmath,amssymb,nofootinbib,11pt]{revtex4} 
\usepackage{graphicx}

\begin{document}

\title{Search for the $a_0(980)-f_0(980)$ mixing in weak decays of $D_s/B_s$ mesons}

\author{Wei Wang$^{1,2}$ }
\affiliation{
$^1$INPAC, Shanghai Key Laboratory for Particle Physics and Cosmology, Department of Physics and Astronomy, Shanghai Jiao-Tong University, Shanghai, 200240,   China\\
$^2$ State Key Laboratory of Theoretical Physics, Institute of Theoretical Physics, Chinese Academy of Sciences, Beijing 100190, China}

\begin{abstract}
Scalar mesons $a^0_0(980)$ and $f_0(980)$ can mix with each other through isospin violating effects, and the mixing intensity has been predicted at the percent level in various theoretical  models.  However the mixing  has not been firmed   established on the experimental side to date.  In this work we explore the possibility to extract  the $a_0-f_0$ mixing intensity  using   weak decays of heavy mesons:  $D_s\to [\pi^0\eta, \pi\pi] e^+\nu$, $B_s\to [\pi^0\eta, \pi\pi]\ell^+\ell^-$ and the $B_s\to J/\psi [\pi^0\eta,\pi^+\pi^-]$ decays.  Based on the large amount of data accumulated by  various experimental facilities including BEPC-II, LHC, Super KEKB and  the future colliders, we find that   the $a_0-f_0$ mixing intensity might  be  determined to a high precision, which will  lead to a better understanding of the nature of scalar mesons. 
\end{abstract}
 
\maketitle

\section{Introduction}

Light scalar mesons below 1GeV play an important role in understanding the QCD vacuum since they share the same quantum numbers $J^{PC}$. But due to the nonperturbative nature of QCD at low energy  the internal structure of scalar mesons is extremely complicated and still under controversy. They 
have been interpreted as quark-antiquark, tetra-quarks,
hadronic molecule, quark-antiquark-gluon hybrid, and etc~\cite{Agashe:2014kda}. 

Among various phenomena,  it is anticipated that  the mixing between the $a^0_0(980)$ and $f_0(980)$
resonances  may shed   light on the
nature of these two resonances, and therefore  has been studied
extensively on   different aspects  and in various processes. For an incomplete list of discussions in the literature, please see Refs.~\cite{Achasov:1979xc,Achasov:1981de,Achasov:1996qn,Krehl:1996rk,Kerbikov:2000pu,Close:2000ah,Kudryavtsev:2001ee,Grishina:2001zj,Close:2001ay,Kudryavtsev:2002uu,Kondratyuk:2002yf,Achasov:2002hg,Achasov:2003se,Grishina:2004rd,Achasov:2004ur,Wu:2007jh,Wu:2008hx,Hanhart:2007bd,Wu:2011yx,Aceti:2012dj,Roca:2012cv,Tarasov:2013yma,Sekihara:2014qxa,Aceti:2015zva} and references therein. 
To date no firm experimental determination on this quantity
is available yet.  The possibility of extracting the $a^0_0(980)$-$f_0(980)$ mixing
from the $J/\psi\to\phi a^0_0(980)\to\phi\eta\pi^0$ reaction has been explored in Refs.~\cite{Wu:2007jh,Wu:2008hx}. This reaction is an isospin breaking process with the initial state of isospin 0 and the
final state of isospin 1. BES-III collaboration has used this process to determine the mixing~\cite{Ablikim:2010aa}: 
\begin{eqnarray}
\xi_{fa}^{J/\psi} \equiv \frac{{\cal B}(J/\psi \to \phi f_0(980)\to \phi a_0^0(980)\to \phi \eta\pi^0)}{ {\cal B}(J/\psi \to \phi f_0(980)\to \phi\pi^+\pi^-)} = (0.60\pm 0.20\pm 0.12\pm 0.26)\%,
\end{eqnarray}
where the uncertainties are statistical, systematics due to the measurement and the parametrization, respectively.  
As one can see, the statistical significance is only about $3.4\sigma$. 

To  more precisely determine the mixing intensity,  two parallel   researches  can be conducted in the future. On the one hand, one may collect more data on the $J/\psi$ (and $\psi'$) and accordingly the errors in this quantity can be reduced significantly. On the other side, one may look for new channels that can be used to determine the mixing parameter.  This will also provide a cross-check of the results derived from the $J/\psi$ decays. 
In this work, we will focus on the latter category.  Weak decays of heavy mesons are not only of great value to determine the standard model parameters (see Ref.~\cite{Wang:2014sba} for a recent review), but can  also provide an ideal platform to study hadron structures~\cite{Oset:2016lyh}.  In the following, we will examine the possibility to extract the mixing intensity from the rare decays of  $D_s$ and $B_s$: $D_s\to [\pi^0\eta, \pi\pi] e^+\nu$, $B_s\to [\pi^0\eta, \pi\pi]\ell^+\ell^-$ and the $B_s\to J/\psi [\pi^0\eta,\pi^+\pi^-]$ decays. An advantage in these modes is that the lepton (or the $J/\psi$) is an iso-singlet system and thus there is a natural isospin filter. At the quark level, the intermediate state  has  $I=0$.  It should be noticed that the semileptonic  $D_s$ and $B_s$ decays into the $\pi^+\pi^-$ via the $f_0(980)$ have already been observed by CLEO-c~\cite{Yelton:2009aa,Ecklund:2009aa,Hietala:2015jqa}  and LHCb collaboration~\cite{Aaij:2014lba}, respectively. The branching fraction of the $B_s\to J/\psi f_0(980)\to  J/\psi\pi^+\pi^-$ is also  measured in Refs.~ \cite{Aaij:2011fx,LHCb:2011ab,LHCb:2012ae,Aaij:2013zpt,Aaij:2014emv,Aaij:2014siy,Li:2011pg,Aaltonen:2011nk,Abazov:2011hv}.

The rest of this paper is  organized as follows. In Sec.\ref{eq:mixing_mechanism}, we will give a brief overview of the $a_0-f_0$ mixing mechanism. We will discuss the mixing effects in $B_s$ and $D_s$ decays in Sec.~\ref{eq:mixing_Bs_Ds}. A short summary is presented in the last section.

\section{The $f_0(980)-a_0(980)$ mixing mechanism}
\label{eq:mixing_mechanism}

\begin{figure}\begin{center}
\includegraphics[scale=0.6]{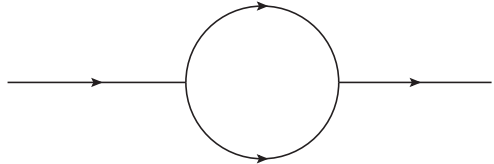}
\caption{One-loop corrections to two-point function.   } \label{fig:mixing_width}
\end{center}
\end{figure}

For the nearly degenerate $a^0_0(980)$ with isospin 1 and $f_0(980)$ with isospin 0, both can  couple to the $K\bar K$ state,  but 
the charged and neutral kaon thresholds are different by about 8 MeV. This difference leads to  the $a^0_0(980)$-$f_0(980)$ mixing.  In the following we will use the abbreviation $a_0$ and $f_0$ to denote the $a_0^0(980)$ and $f_0(980)$ for simplicity.

For illustration, we consider the propagation of the $f_0(980)$ and include the loop corrections through two pseudo-scalars $M_1$ and $M_2$. The  one-loop corrections are  shown in Fig~\ref{fig:mixing_width}.  If one sums these loop corrections in the chain approximation, the $f_0(980)$  propagator will become: 
\begin{eqnarray}
G(s) &\equiv&  \frac{i}{D_f(s)} = \frac{i}{s-m_{f_0}^2} + \frac{i}{s-m_{f_0}^2}  (-i{\cal M}^2) \frac{i}{s-m_{f_0}^2} + ...   \nonumber\\
&=&  \frac{i}{s-m_{f_0}^2 -{\cal M}^2}. 
\end{eqnarray}
with the loop corrections
\begin{eqnarray}
 -i{\cal M}^2&=&  ig_{f_0M_1M_2}  ig_{f_0M_1M_2}^* \int \frac{d^{4}k}{(2\pi)^4}  \frac{i}{k^2-m_{M_1}^2} \frac{i}{(k-p)^2-m_{M_2}^2}.
\end{eqnarray}
Here the $g_{f_0M_1M_2}$ denotes the coupling of the $f_0$ with the $M_1,M_2$. 
The real part of the ${\cal M}^2$  will renormalize the bare mass, leading to the pole in the propagator as the physical mass.  The remanent  multiplicative constant in the real part  is absorbed by the field strength  renormalization factor.  The imaginary part of the ${\cal M}^2$ will result in  a nonzero mass-dependent decay width:
\begin{eqnarray}
 \Gamma^f_{12}(s) = - \frac{1}{\sqrt s}  {\rm Im}[{\cal M}^2] (s) = \frac{1}{16\pi \sqrt s}  |g_{f_0M_1M_2} |^2 \rho_{12}(s),
\end{eqnarray}
with $\rho_{bc}(s)=\sqrt{[1-(m_{b}-m_{c})^{2}/s][1-(m_{b}+m_{c})^{2}/s]}$. 

With the incorporation of the   mixing effects, we have  the $a_0/f_0$ propagator: 
\begin{eqnarray}
G(s)=\frac{i}{D_{f}(s)D_{a}(s)}
\begin{pmatrix}
D_{a}(s)&D_{af}(s)\\D_{af}(s)&D_{f}(s)
\end{pmatrix},
\end{eqnarray}
where $D_{a}$ and $D_{f}$ are the denominators of  the resummed 
propagators for the $a^0_0(980)$ and $f_0(980)$, respectively:
\begin{eqnarray}
D_{a}(s)&=&s-m_{a}^{2}+ i\sqrt{s}[\Gamma^a_{\eta\pi}(s) + \Gamma^a_{K\bar K}(s)],
\\
D_{f}(s)&=&s-m_{f}^{2}+ i\sqrt{s}[\Gamma^f_{\pi\pi}(s)+\Gamma^f_{K\bar K}(s)],
\end{eqnarray}
Since the mixing  term is already small at leading order, it is not necessary to sum all order corrections. We have the expression for the $D_{af}$:
\begin{eqnarray}
D_{af}(s)&=&i \frac{g_{a^0_0(980) K^{+}K^{-}}g_{f_0(980) K^{+}K^{-}}}{16\pi}
\Big\{\rho_{K^{+}K^{-}}(s)-\rho_{K^{0}\bar{K}^0}(s)\Big\}.\label{4}
\end{eqnarray}
The relation between  the $D_{af}$ and the mass-dependent $f_0\to a_0$ mixing parameter
$\xi$ is given as:
\begin{eqnarray} 
\xi(s)=\left|\frac{D_{af}(s)}{D_{a}(s)}\right|^2=\left|\frac{g_{a^0_0(980) K^{+}K^
{-}}g_{f_0(980)
K^{+}K^{-}}[\rho_{K^{+}K^-}(s)-\rho_{K^{0}\bar{K}^0}(s)]}{16\pi
D_{a}(s)}\right|^2.
\end{eqnarray}
As one can see, the mixing parameter arises due to the different masses of the charged and neutral Kaon. The results also  rely on the couplings $g_{a^0_0(980) K^{+}K^{-}}$,  $g_{f_0(980) K^{+}K^{-}}$ and the mass pole position  in the propagator.  
Various  theoretical models predict different values for these quantities, and a thorough  discussion has been presented in Refs.~\cite{Wu:2007jh,Wu:2008hx}.

\section{Mixing effects in the $B_s$ and $D_s$ decays}
\label{eq:mixing_Bs_Ds}

\begin{figure}\begin{center}
\includegraphics[scale=0.7]{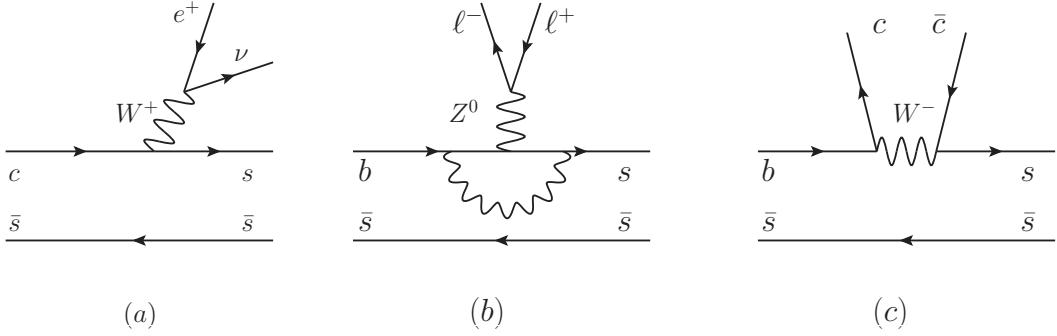}
\caption{Feynman diagrams for  the $D_s$ and $B_s$ decays into the $f_0(980)$ with the $\bar ss$ component  at the quark level. The  panel (a) denotes the semileptonic $D_s$ decay, in which the lepton pair $e^+\nu$ is emitted. One and typical Feynman  diagram for the semileptonic $B_s\to f_0\ell^+\ell^- (\ell=e,\mu,\tau)$ decay are  given in panel (b). The last panel  (c) corresponds to the nonleptonic $B_s$  decay into the $J/\psi$.  } \label{fig:Ds_feynman}
\end{center}
\end{figure}

In this section, we will analyze the mixing intensity in the semileptonic decays of $B_s$ and $D_s$ mesons.  More explicitly, the considered decay processes include
\begin{eqnarray}
 D_s\to \pi^0\eta e^+\nu, \;\;\;\;  D_s\to \pi\pi e^+\nu,\\
 B_s\to \pi^0\eta \ell^+\ell^-, \;\;\;\;  B_s\to \pi\pi  \ell^+\ell^-,\\
 B_s\to \pi^0\eta J/\psi, \;\;\;\;  B_s\to \pi\pi J/\psi. 
\end{eqnarray}

We will  take the $D_s$ decay as the example, whose Feynman diagram is shown in the panel (a) of Fig.~\ref{fig:Ds_feynman}. 
After emitting the off-shell $W$-boson,  the  hadronic sector is the $\bar ss$  which will couple to the iso-singlet component $f_0(980)$. 
Then the decay amplitudes for the $ D_s\to \pi\pi e^+\nu\equiv D_s\to f_0 e^+ \nu\to \pi\pi e^+\nu$  and $ D_s\to \pi^0\eta e^+\nu\equiv D_s\to f_0e^+ \nu\to a_0^0e^+ \nu\to \pi^0\eta e^+ \nu$  are given as
\begin{eqnarray}
{\cal A}( D_s\to  \pi\pi e^+\nu)   &=& \hat A \left\{   \frac{i}{D_{f_0}} \times  ig_{f_0\pi\pi} \right\}, \nonumber\\
{\cal A}( D_s\to  \pi^0\eta e^+\nu) &=& \hat A  \left\{ \frac{i}{D_{f_0} D_{a}} D_{fa}  \times  ig_{a_0\pi\eta} \right\},
\end{eqnarray}
where  the amplitude $\hat A$ can be expressed in terms of the transition form factors:
\begin{eqnarray}
&&\langle f_0(p_{f_0})|\bar s \gamma_\mu\gamma_5 c | {D}_s(p_{D_s})\rangle =  -i \Big\{F_1(q^2)\Big[P_\mu
 -\frac{m_{D_s}^2-m_{f_0}^2}{q^2}q_\mu\Big] + F_0(q^2)\frac{m_{D_s}^2-m_{f_0}^2}{q^2}q_\mu\Big\}, 
\end{eqnarray}
The double differential decay width is then derived as
\begin{eqnarray}
 \frac{d\Gamma(D_s\to \pi\pi e^+\nu)}{ds dq^2}  =  \frac{ \lambda^{3/2} G_F^2 |V_{cs}|^2}{192 m_{D_s}^3\pi^3}  F_1^2(q^2).   \frac{\sqrt{s}}{\pi |D_{f}(s)|^2} \Gamma^f_{\pi\pi}(s),\\
 \frac{d\Gamma(D_s\to \pi^0\eta e^+\nu)}{ds dq^2}  =  \frac{\lambda^{3/2}G_F^2 |V_{cs}|^2}{192 m_{D_s}^3\pi^3}  F_1^2(q^2).   \frac{\sqrt{s} |D_{af}(s)|^2}{\pi |D_{f}(s)D_a(s)|^2} \Gamma^a_{\pi\eta}(s),
\end{eqnarray}
where $q^2$ is the invariant mass of the lepton pair, and the $s$ is the invariant mass square of the two pseudo-scalars.  
Here $G_F$ is the Fermi constant,  $V_{cs}$ is the CKM matrix element, and the   K\"allen function $\lambda$ is: $ \lambda = m_{D_s}^4+ s^2 + (q^2)^2- 2(m_{D_s}^2 q^2+m_{D_s}^2 s+s q^2)$.

Since in this work we are interested in the  mixing intensity  in the $a_0(980)-f_0(980)$ resonance  region, one may integrate out the $q^2$ first, leading to
\begin{eqnarray} 
\frac{d\Gamma( D_s\to  \pi\pi e^+\nu) }{ds}  = C  \frac{\sqrt{s}}{\pi |D_{f}(s)|^2} \Gamma^f_{\pi\pi}(s),\\
\frac{d\Gamma(D_s\to\pi^0\eta e^+\nu)}{ds }  = C  \frac{\sqrt{s} |D_{fa}(s)|^2}{\pi |D_{f}(s) D_a(s) |^2} \Gamma^a_{\pi\eta}(s),
\end{eqnarray}
where the coefficient $C$ is obtained via the integration over $q^2$. 
The mass-dependent  mixing intensity can be defined as 
\begin{eqnarray}
 \xi_{fa}^{D_s}(s)  &\equiv& \frac{d \Gamma(D_s\to \pi^0\eta e^+\nu)/ds}{ d\Gamma(D_s\to \pi\pi e^+\nu)/ds} \nonumber\\
 &=&   \frac{|D_{af}(s)|^2  \Gamma^a_{\pi\eta}(s) }{ |D_{a}(s)|^2 \Gamma^{f}_{\pi\pi}(s)},
\end{eqnarray}
while in experiments   one can directly measure the integrated mixing intensity: 
\begin{eqnarray}
\overline  \xi_{fa}^{D_s}   &\equiv& \frac{  \Gamma (D_s\to   \pi^0\eta e^+\nu)}{  \Gamma (D_s\to \pi\pi e^+\nu)}\nonumber\\
&\equiv&\frac{\int_{s'_{min}}^{s'_{max}}ds d \Gamma(D_s\to \pi^0\eta e^+\nu)/ds}{ \int_{s_{min}}^{s_{max}} ds d\Gamma(D_s\to \pi\pi e^+\nu)/ds}\nonumber\\
 &=&      \int_{s'_{min}}^{s'_{max}} ds \frac{\sqrt{s} |D_{fa}(s)|^2}{  |D_{f}(s) D_a(s) |^2} \Gamma^a_{\pi\eta}(s) \bigg/ \int_{s_{min}}^{s_{max}} ds \frac{\sqrt{s}}{ |D_{f}(s)|^2} \Gamma^f_{\pi\pi}(s).  \label{eq:averaged_mixing_intensity}
\end{eqnarray}
Here the $s^{(')}_{min}$ and $s^{(')}_{max}$ denotes the lower and upper invariant mass cuts.  In the previous  BES-III analysis of the mixing intensity using the $J/\psi$ decays~\cite{Ablikim:2010aa}, the mass of the mixing signal is set to 991.3 MeV at the center of charged and neutral kaon thresholds, and the width of the mixing signal is set to 8 MeV. It corresponds to
\begin{eqnarray}
 s'_{min}= [(991.3-4){\rm MeV}]^2,\;\;\;  s'_{max}= [(991.3+4){\rm MeV}]^2. \label{eq:pieta_cuts}
\end{eqnarray}
For the $f_0(980)$, one may follow the  BES-III analysis  of the $J/\psi\to\phi\pi^+\pi^-$~\cite{Ablikim:2004wn}:
\begin{eqnarray}
 s_{min} = [900{\rm MeV}]^2, \;\;\;
 s_{max} = [1000{\rm MeV}]^2.  \label{eq:pipi_cuts}
\end{eqnarray}

With the meson masses (in units of MeV) taken from Particle Data Group~\cite{Agashe:2014kda}
\begin{eqnarray}
 m_{K^+}= 493.677, \;\;\; m_{K^0}= 497.614,\;\;\; m_{\pi^0}= 134.9766,\;\;\; m_{\eta}= 547.862,
\end{eqnarray}
we update the predictions for the mixing intensity $\xi_{fa}(s)$ at $\sqrt{s}=991.3$ MeV  and give the  results for the integrated quantity $\overline  \xi_{fa}$ with the kinematics in Eqs.~(\ref{eq:pieta_cuts}) and (\ref{eq:pipi_cuts})  in table~\ref{tab:mixing_values}. In the calculation, the isospin symmetry has been used for the $\pi\pi$ sytem.  Results for the $\xi_{fa}$ are consistent with Refs.~\cite{Wu:2007jh,Wu:2008hx}.  As one can see from this table, most predictions for the integrated mixing intensity are at the percent level.

\begin{table}[ht]
\caption{Meson masses (in units of MeV) and couplings (in units of GeV) predicted by various models or determined by experimental measurements. The mixing intensity $\xi_{fa}(s)$ (in unit  of $\%$) is evaluated at $\sqrt{s}=991.3$ MeV, which is at the center the $K^+K^-$ and $K^0\bar K^0$ threshold. The integrated  mixing intensity $\bar \xi_{fa}$ (in unit  of $\%$) is evaluated by Eq.~(\ref{eq:averaged_mixing_intensity}) with the kinematics in Eqs.~(\ref{eq:pieta_cuts}) and (\ref{eq:pipi_cuts}).  }\label{tab:mixing_values}
\begin{tabular}{|c|c|c|c|c|c|c|c|c|c|}
\hline   model/experiment & $m_{a_0}$ & $g_{a_{0}\pi\eta}$ &
$g_{a_{0}K^{+}K^{-}}$ & $m_{f_0}$ & $g_{f_{0}\pi^{0}\pi^{0}}$ & $g_{f_{0}K^{+}K^{-}}$ & $\xi_{fa}(\%)$ & $\overline \xi_{fa}(\%)$\\
\hline
   $q\bar{q}$ model \cite{Achasov:1987ts} & 983 & 2.03 & 1.27 & 975 & 0.64 & 1.80 & 2.2 & 0.6  \\
\hline
  $q^{2}\bar{q}^{2}$ model \cite{Achasov:1987ts} & 983 & 4.57 & 5.37& 975 & 1.90 & 5.37 & 6.5 & 1.2  \\
\hline
 $K\bar{K}$ model \cite{Achasov:1997ih,Weinstein:1983gd,Weinstein:1990gu} & 980 & 1.74 & 2.74& 980 & 0.65 & 2.74 & 20.1 & 7.5   \\
\hline
 $q\bar{q}g$ model \cite{Ishida:1995} & 980 & 2.52& 1.97& 975 & 1.54 & 1.70 & 0.5  & 0.05\\
\hline
 SND \cite{Achasov:2000ym,Achasov:2000ku} & 995 & 3.11 & 4.20& 969.8 & 1.84 & 5.57 & 8.5 & 1.7  \\
\hline
  KLOE \cite{Aloisio:2002bsa,Aloisio:2002bt} & 984.8 & 3.02 & 2.24& 973 & 2.09 & 5.92 & 3.2 & 0.6   \\
\hline
  BNL~\cite{Teige:1996fi,Zou:1993az}& 1001  & 2.47 & 1.67 & 953.5 & 1.36 & 3.26  & 1.8 & 0.4  \\
\hline
 CB~\cite{Bugg:1994mg,Ablikim:2004wn} & 999 & 3.33 & 2.54  & 965  & 1.66  & 4.18  & 2.6 & 0.5  \\
\hline
\end{tabular}
\end{table}

The CLEO collaboration has firstly measured the branching fraction~\cite{Ecklund:2009aa}: 
\begin{eqnarray}
{\cal B}(D_s\to f_0(980)(\to \pi^+\pi^-) e^+ \nu_{e})= (2.0\pm0.3\pm0.1)\times 10^{-3}, \label{eq:Ds_f0_pipi_CLEO_c}
\end{eqnarray}
but a recent analysis based on the CLEO-c data gives a similar result with a smaller central value~\cite{Hietala:2015jqa}: 
\begin{eqnarray}
{\cal B}(D_s\to f_0(980)(\to \pi^+\pi^-) e^+ \nu_{e})= (1.3\pm0.2\pm0.1)\times 10^{-3}. \label{eq:Ds_f0_pipi_CLEO_c_2}
\end{eqnarray}
In  near future the BES-III collaboration  will collect about $3fb^{-1}$ data in $e^+e^-$ collision  at the energy around $4.18$GeV~\cite{Asner:2008nq}. This corresponds to a few times $10^6$ events of the $D_s$ mesons and  accordingly a few thousand events for the $D_s\to \pi^+\pi^- e^+\nu$ decay before any kinematics cut. As we can see if the mixing intensity is at the percent level, there is a promising  prospect to measure/constrain the mixing by  BES-III collaboration using the $D_s\to [\pi^0\eta,\pi^+\pi^-] e^+\nu$.

The analysis of the $B_s\to [\pi^0\eta, \pi\pi]\ell^+\ell^-$   and $B_s\to [\pi^0\eta, \pi\pi]J/\psi$ (with $\ell=e,\mu,\tau$)  is also similar.  For instance in the semileptonic decay,  one can study the mass-dependent and integrated mixing intensity which is defined as 
\begin{eqnarray}
 \xi_{fa}^{B_s}(s)  &\equiv& \frac{d \Gamma(B_s\to \pi^0\eta \ell^+\ell^-)/ds}{ d\Gamma(B_s\to \pi\pi \ell^+\ell^-)/ds} \nonumber\\
 &=&   \frac{|D_{af}(s)|^2  \Gamma^a_{\pi\eta}(s) }{ |D_{a}(s)|^2 \Gamma^{f}_{\pi\pi}(s)},\\
\overline  \xi_{fa}^{B_s}     &\equiv& \frac{  \Gamma (B_s\to \pi^0\eta \ell^+\ell^-)}{  \Gamma (B_s\to \pi\eta \ell^+\ell^-)} 
\nonumber\\
 &=&   \int_{s'_{min}}^{s'_{max}} ds \frac{\sqrt{s} |D_{fa}(s)|^2}{  |D_{f}(s) D_a(s) |^2} \Gamma^a_{\pi\eta}(s) \bigg/ \int_{s_{min}}^{s_{max}} ds \frac{\sqrt{s}}{ |D_{f}(s)|^2} \Gamma^f_{\pi\pi}(s). 
\end{eqnarray}
For the rare decay $B_s\to f_0(\to \pi^+\pi^-)\mu^+\mu^-$, the LHCb collaboration  has performed a detailed  analysis with the result~\cite{Aaij:2014lba}: 
\begin{eqnarray}
{\cal B}(B_s\to f_0(980)(\to \pi^+\pi^-) \mu^+\mu^-) = (8.3\pm 1.7)\times 10^{-8}.  \label{eq:Bs_f0_mumu_data}
\end{eqnarray}    
This has already triggered some theoretical interpretations  using  two-meson light-cone distribution amplitudes (LCDAs)~\cite{Wang:2015uea,Wang:2015paa}. 
The LHCb collaboration has also systematically studied  the $B_s\to J/\psi\pi^+\pi^-$ decays~\cite{Aaij:2011fx,LHCb:2011ab,LHCb:2012ae,Aaij:2013zpt,Aaij:2014emv,Aaij:2014siy}, and some implications on the structure of scalar mesons have been explored in Refs.~\cite{Bayar:2014qha,Close:2015rza,Sekihara:2015iha}. The averaged branching fraction is  given as~\cite{Agashe:2014kda}
\begin{eqnarray}
{\cal B}(B_s\to J/\psi f_0(980)(\to \pi^+\pi^-)) =(1.35\pm 0.16)\times 10^{-4}. 
\end{eqnarray}  
Since much more data will be collected by  experimental facilities including the LHCb detector~\cite{Bediaga:2012py}  the Super-B factory at the KEK~\cite{Aushev:2010bq}, it is likely to precisely  derive the $a_0(980)$ and $f_0(980)$ mixing from these  weak decays of heavy mesons.

\section{Summary }

To understand the internal structure of  light scalar mesons is a long-standing problem in hadron physics.  It is expected that some aspects can be unraveled by the study of $a^0_0(980)-f_0(980)$ mixing.  The two scalar mesons can couple to the $K-\bar K$ and will  mix with each other due to the different masses for the charged and neutral kaons. The mixing intensity has been predicted at the percent level in various theoretical  models.   A number of processes have been proposed to study the mixing, but 
to date there is no firm evidence  on the experimental side. 

In this work we have proposed to use the weak decays of the $B_s$ and $D_s$ mesons to study the $a_0-f_0$ mixing. We have studied the semileptonic decays of heavy mesons,  $D_s\to [\pi^0\eta, \pi\pi] e^+\nu$, $B_s\to [\pi^0\eta, \pi\pi]\ell^+\ell^-$ and the $B_s\to J/\psi [\pi^0\eta,\pi^+\pi^-]$ decays.  Based on the large amount of data accumulated by  various experimental facilities including BEPC-II, Super KEKB, LHC and  the future colliders like the High Intensity Electron Positron Accelerator (HIEPA) expected  running at $2-7$ GeV with the designed luminosity of $10^{35}cm^{-2}s^{-1}$, the  Z-factory running at $Z$-pole  and the circular electron-positron collider (CEPC), it is very likely that  the $a_0-f_0$ mixing intensity  can  be  determined to a high precision, which will  lead to a better  understanding of the nature of scalar mesons.

\section*{Acknowledgements}
The author  is very grateful to Jian-Ping Dai,  Liao-Yuan Dong, Hai-Bo  Li, Cai-Dian L\"u,  Jia-Jun Wu,   Lei Zhang  and Qiang Zhao for  enlightening discussions. This work was supported in part  by National  Natural  Science Foundation of China under Grant  No.11575110,  Natural  Science Foundation of Shanghai under Grant  No. 15DZ2272100 and No. 15ZR1423100,  by the Open Project Program of State Key Laboratory of Theoretical Physics, Institute of Theoretical Physics, Chinese  Academy of Sciences, China (No.Y5KF111CJ1), and  by   Scientific Research Foundation for   Returned Overseas Chinese Scholars, State Education Ministry.



\end{document}